\documentclass[journal]{IEEEtran}
\usepackage{amsfonts}
\usepackage{cite,graphicx,amsmath,amsthm}
\allowdisplaybreaks[4]
\usepackage{subfigure}
\usepackage{fancyhdr}
\usepackage{dsfont}
\usepackage{array,color}
\usepackage{bm}
\usepackage{booktabs}
\usepackage{multirow}
\usepackage{algorithm}
\usepackage{algpseudocode}

\usepackage{bbm}
\usepackage{graphicx}

\newtheorem{theorem}{Theorem}

\newtheorem{lemma}{Lemma}

\newtheorem{proposition}{Proposition}

\newtheorem{corollary}{Corollary}

\newtheorem{property}{Property}

\newtheorem{remark}{Remark}

\newtheorem{claim}{Claim}

\begin{document}
\title{{\huge Learning Centric Power Allocation for Edge Intelligence}}

\author{Shuai Wang$^{*\diamond}$, Rui Wang$^{*}$, Qi Hao$^{\diamond}$, Yik-Chung Wu$^{\dag}$, and H.~Vincent~Poor$^{\star}$\\
        $^{*}$Department of Electrical and Electronic Engineering, $^{\diamond}$Department of Computer Science and Engineering,\\$^{*\diamond}$Southern University of Science and Technology, Shenzhen 518055, China\\
        $^{\dag}$Department of Electrical and Electronic Engineering, The University of Hong Kong, Hong Kong\\
        $^{\star}$Department of Electrical Engineering, Princeton University, Princeton, NJ 08544 USA \\
E-mail: \{wangs3, wang.r, hao.q\}@sustech.edu.cn; ycwu@eee.hku.hk; poor@princeton.edu
}

\maketitle
\begin{abstract}
While machine-type communication (MTC) devices generate massive data, they often cannot process this data due to limited energy and computation power.
To this end, edge intelligence has been proposed, which collects distributed data and performs machine learning at the edge.
However, this paradigm needs to maximize the learning performance instead of the communication throughput, for which the celebrated water-filling and max-min fairness algorithms become inefficient since they allocate resources merely according to the quality of wireless channels.
This paper proposes a learning centric power allocation (LCPA) method, which allocates radio resources based on an empirical classification error model.
To get insights into LCPA, an asymptotic optimal solution is derived.
The solution shows that the transmit powers are inversely proportional to the channel gain, and scale exponentially with the learning parameters.
Experimental results show that the proposed LCPA algorithm significantly outperforms other power allocation algorithms.
\end{abstract}


\section{Introduction}

Machine learning is revolutionizing every branch of science and technology \cite{dl1}.
If a machine wants to learn, it requires at least two ingredients: information and computation, which are usually separated from each other in machine-type communication (MTC) systems \cite{iot1}.
To address this challenge brought by MTC, a promising solution is the \emph{edge intelligence} technique \cite{edge1,edge2,edge3,edge4,edge5} that uses an intelligent edge to collect data generated from MTC devices and trains a machine learning model or \emph{fine-tunes a pre-trained model} at the edge.

In contrast to conventional communication systems, edge intelligence systems aim to maximize the learning performance instead of the communication throughput.
Therefore, edge intelligence resource allocation becomes very different from traditional resource allocation schemes that merely consider the wireless channel conditions \cite{waterfilling,fair,sumrate}.
For instance, the celebrated water-filling scheme allocates more resources to better channels for throughput maximization \cite{waterfilling}, and the max-min fairness scheme allocates more resources to cell-edge users to maintain certain quality of service \cite{fair}.
While these two schemes have proven to be very efficient in traditional wireless communication systems, they could lead to poor learning performance in edge intelligence systems, because they do not account for the machine learning factors such as model and dataset complexities.
Imagine training a deep neural network (DNN) and a support vector machine (SVM) at the edge.
Due to much larger number of parameters in DNN, the edge should allocate more resources to MTC devices that upload data for the DNN than those for the SVM.

Nonetheless, in order to maximize the learning performance in the resource allocation, we need a mathematical expression of the learning performance with respect to the data size, which does not exist to the best of the authors' knowledge.
Fortunately, it has been proved in \cite{model1} that the learning performance (i.e., the generalization error) can always be upper bounded by the summation of the bias between the main prediction and the optimal prediction, the variance due to training datasets, and the noise of the target example.
Moreover, for certain loss functions (e.g., squared loss and zero-one loss), the bound is tight.
Based on this bias-variance decomposition theory, an empirical nonlinear classification error model has been proposed in \cite{model3,model2,model5}, with parameters obtained from curve fitting of experimental data.
The model is also theoretically supported by the asymptotic analysis based on statistical mechanics \cite{model4}.

In this paper, we adopt the above nonlinear model to approximate the learning performance, and a \emph{learning centric power allocation (LCPA)} problem is formulated with the aim of minimizing classification error subject to the total power budget constraint.
By leveraging the majorization minimization (MM) framework from optimization, the LCPA algorithm that converges to a Karush-Kuhn-Tucker (KKT) solution is proposed.
To get deeper insights into LCPA, an analytical solution is derived for the asymptotic case, where massive multiple-input multiple-output (MIMO) technique is employed at the edge.
The asymptotic optimal solution discloses that the transmit powers are inversely proportional to the channel gain, and scale exponentially with the classification error model parameters.
This result reveals that machine learning has a stronger impact than wireless channels in LCPA.
Experimental results based on public datasets show that the proposed LCPA is able to achieve a higher classification accuracy than that of the sum-rate maximization and max-min fairness power allocation schemes.
For the first time, the benefit brought of joint communication and learning design is quantitatively demonstrated in edge intelligence systems.

\begin{figure*}[t!]
\centering
\includegraphics[width=170mm]{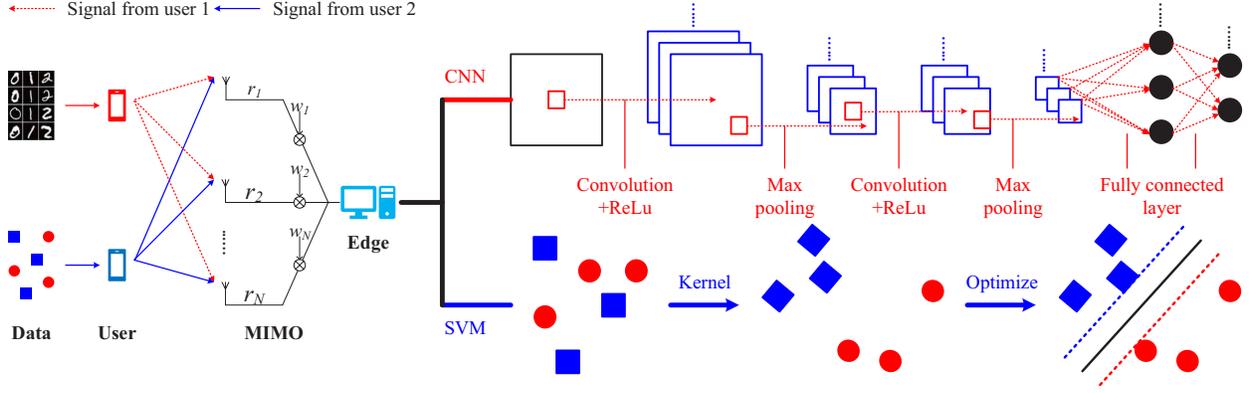}
\caption{System model of edge intelligence with two users.}
\label{fig_sim}
\end{figure*}

\section{System Model and Problem Formulation}

\setcounter{secnumdepth}{4}
We consider an edge intelligence system, which consists of an intelligent edge (i.e., a radio access point with computation power) with $N$ antennas and $K$ users with datasets $\{\mathcal{D}_1,\cdots,\mathcal{D}_K\}$.
The goal of the edge is to train $K$ classification models by collecting $\{\mathcal{D}_1,\cdots,\mathcal{D}_K\}$ from the $K$ users (e.g., UAVs with cameras), where $\mathcal{D}_k$ is observed at user $k$ and used for training model $k$.
For the classification models, without loss of generality, Fig.~1 depicts the case of $K=2$ with a convolutional neural network (CNN) and a support vector machine (SVM), but more users and other classification models are equally valid.
It is assumed that the data are labeled at the edge.
This can be supplemented by the recent self-labeled techniques \cite{self}, where a classifier is trained with an initial small number of labeled examples with manual labeling, and then the model is retrained with its own most confident predictions, thus enlarging its labeled training set.
After training the classifiers, the edge can feedback the trained models to users for subsequent use (e.g., object recognition).
Notice that if the classifiers are pre-trained at the cloud center and deployed at the edge, the task of edge intelligence is to fine-tune the pre-trained models at the edge, using local and proprietary data generated from MTC users.

More specifically, the user $k\in\{1,\cdots,K\}$ transmits a signal $s_{k}$ with power $\mathbb{E}[|s_{k}|^2]=p_{k}$ for all $k$.
Accordingly, the received signal $\mathbf{r}=[r_1,\cdots,r_N]^T\in\mathbb{C}^{N\times 1}$ at the edge is
$\mathbf{r}=\sum_{k=1}^K\mathbf{h}_{k}\, s_{k}+\mathbf{n}$,
where $\mathbf{h}_{k}\in \mathbb{C}^{N\times 1}$ is the channel vector from the user $k$ to the edge,
and $\mathbf{n}\sim \mathcal{CN}(\mathbf{0},\sigma^2\mathbf{I}_N)$.
By applying the well-known maximal ratio combining (MRC) receiver $\mathbf{w}_k=\mathbf{h}_{k}/||\mathbf{h}_{k}||_2$ to $\mathbf{r}$, the data-rate of user $k$ is
\begin{align}
&R_{k}=\mathrm{log}_2\left(1+\frac{G_{k,k}p_{k}}{\sum_{l=1,l\neq k}^KG_{k,l}p_{l}+
\sigma^2} \right), \label{Rk}
\end{align}
where $G_{k,l}$ represents the composite channel gain (including channel fading and MIMO processing):
\begin{align}
&G_{k,l}=
\left\{
\begin{aligned}
&||\mathbf{h}_{k}||_2^2
,\quad{}&\mathrm{if}~k=l
\\
&\frac{|\mathbf{h}_k^H\mathbf{h}_{l}|^2}{||\mathbf{h}_{k}||_2^2}
,\quad{}&\mathrm{if}~k\neq l
\end{aligned}
\right.
.
\end{align}
With the expression of $R_k$ in \eqref{Rk}, the amount of data in $\mathrm{bit}$ received from user $k$ is $BTR_{k}$, where constant $B$ is the bandwidth in $\mathrm{Hz}$ that is assigned to the system (e.g., a standard MTC system would have $180\,\mathrm{kHz}$ bandwidth \cite{iot3}), and $T$ is the total number of transmission time in second.
As a result, the total number of training samples that are collected at the edge for training the model $k$ is
\begin{align}\label{sample size}
&v_k= \left\lfloor \frac{BTR_{k}}{D_k} \right\rfloor +A_k \approx \frac{BTR_{k}}{D_k} +A_k,
\end{align}
where $A_k$ is the initial number of samples for task $k$ at the edge, $\lfloor x\rfloor=\mathrm{max}\{n\in\mathbb{Z}:n\leq x \}$ and the approximation is due to $\lfloor x\rfloor\to x$ when $x\gg 1$.
Notice that $D_k$ is the number of bits for each data sample in $\mathcal{D}_k$.
For example, the handwritten digits in the MNIST dataset \cite{MNIST} are grayscale images with $28\times 28$ pixels (each pixel has $8$ bits), and in this case $D_k=8\times28\times28+4=6276\,\mathrm{bits}$ ($4$ bits are reserved for the labels of $10$ classes \cite{MNIST} in case the users also transmit labels).
With the collected samples, the intelligent edge can then train its models $\{1,\cdots,K\}$ in the learning phase.
We use the function $\Psi_k(v_k)$ to denote the classification error of the learning model $k$ when the sample size is $v_k$.

In the considered system, the design variables that can be controlled are the transmit powers of different users $\mathbf{p}=[p_1,\cdots,p_K]^T$ and the sample sizes of different models $\mathbf{v}=[v_1,\cdots,v_K]^T$.
Since the power costs at users should not exceed the total budget $P_{\mathrm{sum}}$, the variable $\mathbf{p}$ needs to satisfy
$\sum_{k=1}^Kp_k=P_{\mathrm{sum}}$.
Having the transmit power satisfied, it is then crucial to minimize the classification errors (i.e., the number of incorrect predictions divided by the number of total predictions),
which leads to the following learning centric power allocation (LCPA) problem:
\begin{subequations}
\begin{align}
\mathrm{P}:\mathop{\mathrm{min}}_{\substack{\mathbf{p},\,\mathbf{v}}}
\quad&\mathop{\mathrm{max}}_{k=1,\cdots,K}~\Psi_k(v_k),   \nonumber\\
\mathrm{s. t.}\quad &\sum_{k=1}^Kp_k=P_{\mathrm{sum}},\quad p_k\geq 0,\quad k=1,\cdots,K,  \\
&\frac{BT}{D_k}\mathrm{log}_2\left(1+\frac{G_{k,k}p_{k}}{\sum_{l=1,l\neq k}^KG_{k,l}p_{l}+
\sigma^2} \right)
\nonumber\\
&
+A_k=v_k,\quad k=1,\cdots,K,
 \label{P0b}
\end{align}
\end{subequations}
where the min-max operation at the objective function is to guarantee the worst-case learning performance.
Notice that when each user has its own maximum transmit power, the per-user power constraints $\{p_k\leq P_{\mathrm{max}},\,\forall k\}$ can be added to $\mathrm{P}$, and the LCPA algorithm is still applicable to the resultant problem.

\section{ Classification Error Modeling}

The key challenge to solve $\rm{P}$ is that functions $(\Psi_1,\cdots,\Psi_K)$ are unknown, and to the best of the authors' knowledge, currently there is no exact expression of $\Psi_k(v_k)$.
To address this issue, we will adopt an empirical classification error model to approximate $\Psi_k$.

In general, the classification error $\Psi_k(v_k)$ is a nonlinear function of $v_m$.
Particularly, this nonlinear function should satisfy the following properties:
\begin{itemize}
\item[(i)] Since $\Psi_k$ is a percentage, $0\leq \Psi_k(v_k) \leq 1$;

\item[(ii)] Since more data would provide more information, $\Psi_k(v_k)$ is a monotonically decreasing function of $v_k$ \cite{model2};

\item[(iii)] As $v_k$ increases, the magnitude of derivative $|\partial \Psi_k/\partial v_k|$ would gradually decrease and become zero when $v_k$ is sufficiently large \cite{model3}, meaning that increasing sample size no longer helps machine learning.

\end{itemize}
Based on the properties (i)--(iii), the following nonlinear model $\Theta_k(v_k,|a_k,b_k)$ \cite{model2,model3,model4,model5} can be used to capture the shape of $\Psi_k(v_k)$:
\begin{align}
\Psi_k(v_k) \approx
\Theta_k(v_k,|a_k,b_k)
=a_k\times v_k^{-b_k}, \label{model1}
\end{align}
where $a_k,b_k\geq 0$ are tuning parameters.
The model \eqref{model1} indicates that there is an inverse power relationship between learning performance and the amount of training data \cite{model3,model4,model5}.
It can be seen that $\Theta_k$ satisfies all the features (i)--(iii).
Moreover, $\Theta_k(v_k,|a_k,b_k)\to 0$ if $v_k\to +\infty$, meaning that the error is $0$ with infinite data\footnote{We assume the model is powerful enough such that given infinity amount of data, the error rate can be driven to zero.}.

\textbf{Interpretation from Learning Theory.}
Apart from features (i)--(iii), the error model in \eqref{model1} can also be explained by the bias-variance decomposition theory \cite{model1}.
In particular, it is known that the probability of incorrect classification is proportional to the summation of a bias term and a variance term \cite{model1}.
The bias is independent of the training set, and is zero for a learner that always makes the optimal prediction \cite{model1}.
The variance is independent of the true value of the predicted variable, and is asymptotically proportional to $1/v_k$ for independent and identically distributed (IID) samples \cite{model2}.
But since the datasets could be non-IID, we use $v_k^{-b_k}$ to represent the error rate, with $b_k$ being a tuning parameter to account for the dataset distribution.
Finally, by multiplying a weighting factor $a_k$ to account for the model complexity of the classifier $k$, we immediately obtain the result in \eqref{model1}.

\subsection{Parameter Fitting of CNN and SVM Classifiers}

We use the public MNIST dataset \cite{MNIST} as the input images, and train the $6$-layer CNN (shown in Fig.~1) with training sample size $v_k^{(i)}$ ranging from $100$ to $10000$.
In particular, the input image is sequentially fed into a $5\times 5$ convolution layer (with ReLu activation, 32 channels, and SAME padding), a $2\times 2$ max pooling layer, then another $5\times 5$ convolution layer (with ReLu activation, 64 channels, and SAME padding), a $2\times 2$ max pooling layer, a fully connected layer with $128$ units (with ReLu activation), and a final softmax output layer (with $10$ outputs).
The training procedure is implemented via Adam optimizer with a learning rate of $10^{-4}$ and a mini-batch size of $100$.
After training for $5000$ iterations, we test the trained model on a validation dataset with $1000$ unseen samples, and compute the corresponding classification error.
By varying the sample size $v_k$ as $(v_k^{(1)},v_k^{(2)},\cdots)=(100,150,200,300,500,1000,5000,10000)$, we can obtain the classification error $\Psi_k(v_k^{(i)})$ for each sample size $v_k^{(i)}$, where $i=1,\cdots,Q$, and $Q=8$ is the number of points to be fitted.
With $\{v_k^{(i)},\Psi_k(v_k^{(i)})\}_{i=1}^Q$, the parameters $(a_k,b_k)$ in $\Theta_k$ can be found via the following nonlinear least squares fitting:
\begin{align}
\mathop{\mathrm{min}}_{a_k,\,b_k}\quad~&\frac{1}{Q}\mathop{\sum}_{i=1}^Q\Big|\Psi_k\left(v_k^{(i)}\right)-\Theta_k\left(v_k^{(i)},a_k,b_k\right)\Big|^2,
\nonumber\\
\mathrm{s.t.}\quad\quad&
a_k\geq 0,\quad b_k\geq 0.
\label{fitting}
\end{align}
The above problem is solved by brute-force search.

To demonstrate the versatility of the model, we also fit the nonlinear model to the classification error of a support vector machine (SVM) classifier.
The SVM uses penalty coefficient equal to $1$ and Gaussian kernel function $K(\mathbf{x}_i,\mathbf{x}_j)=\mathrm{exp}\left(-\widetilde{\gamma}\times ||\mathbf{x}_i-\mathbf{x}_j||_2^2\right)$ with $\widetilde{\gamma}=0.001$ \cite{sklearn}.
Moreover, the SVM classifier is trained on the digits dataset in the Scikit-learn Python machine learning tookbox, and the dataset contains $1797$ images of size $8\times 8$ from $10$ classes, with $5$ bits (corresponding to integers $0$ to $16$) for each pixel \cite{sklearn}.
Therefore, each image needs $D_k=8\times 8\times 5+4=324\,\mathrm{bits}$.
Out of all images, we train the SVM using the first $1000$ samples with sample size $(v_k^{(1)},v_k^{(2)},\cdots)=(30,50,100,200,300,400,500,1000)$,
and use the latter $797$ samples for testing.
The parameters $(a_k,b_k)$ for the SVM are obtained following a similar procedure in \eqref{fitting}.

To evaluate the fitted models for SVM and CNN, Fig.~2a illustrates the classification error versus the sample size.
It is observed from Fig.~2a that with the parameters $(a_k,b_k)=(9.27,0.74)$, the nonlinear classification error model in \eqref{model1} matches the experimental data of CNN very well.
On the other hand, with $(a_k,b_k)=(6.94,0.8)$, the model in \eqref{model1} also fits the experimental data of SVM.

\begin{figure*}[htbp]
 \centering
  \subfigure[]{
    \label{fig:subfig:a} 
    \includegraphics[height=45mm]{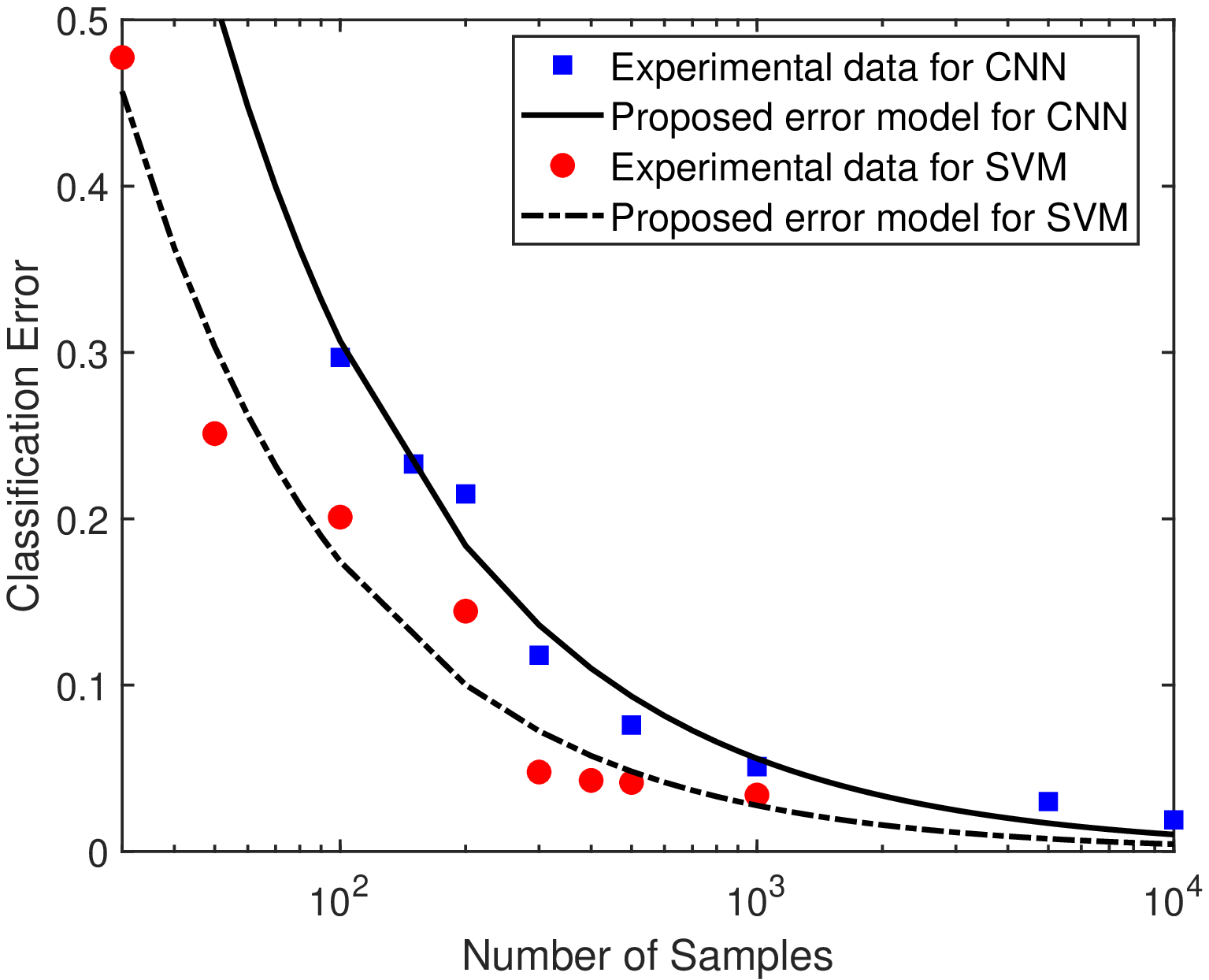}}
  \subfigure[]{
    \label{fig:subfig:a} 
    \includegraphics[height=45mm]{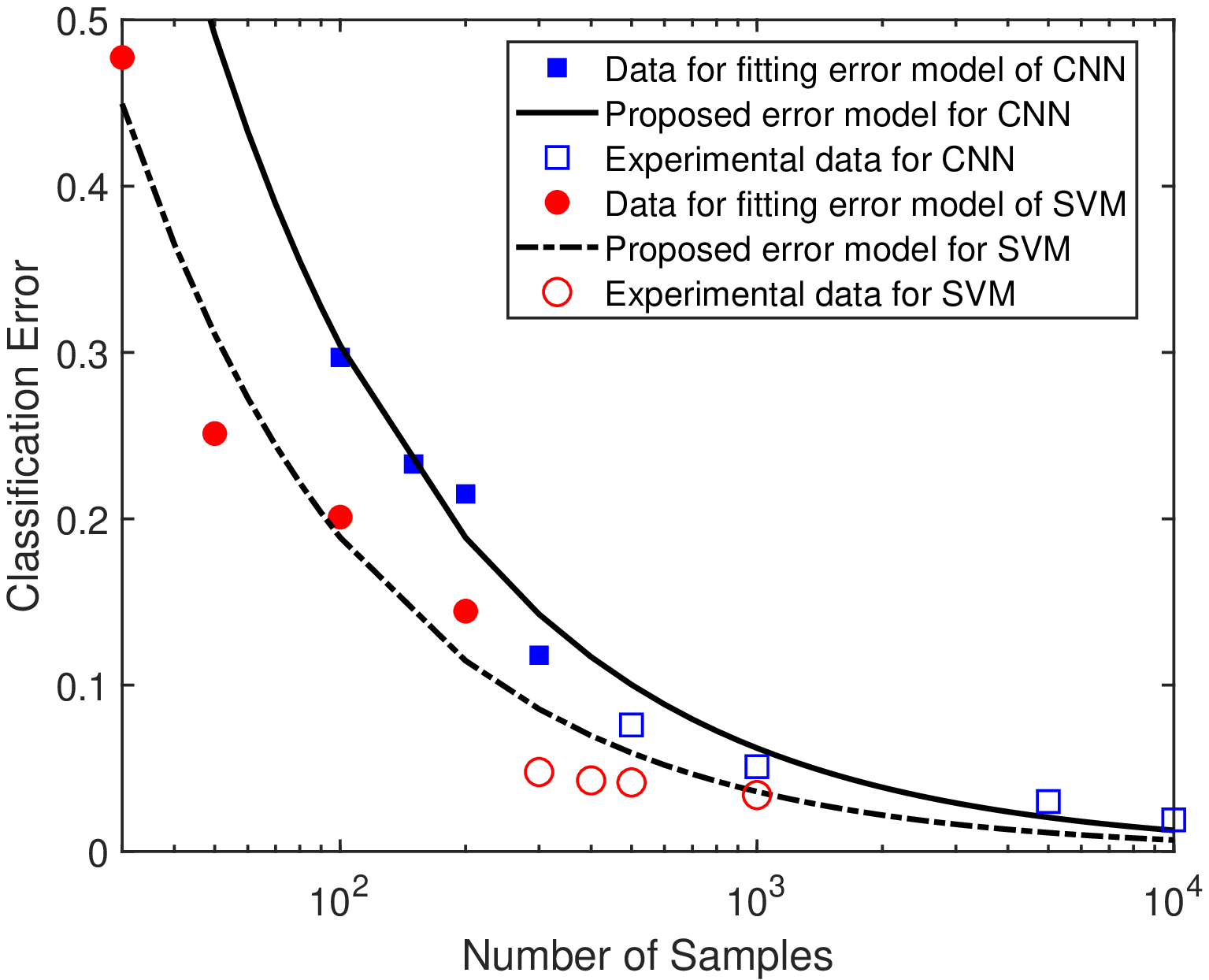}}
  \subfigure[]{
    \label{fig:subfig:b} 
    \includegraphics[height=45mm]{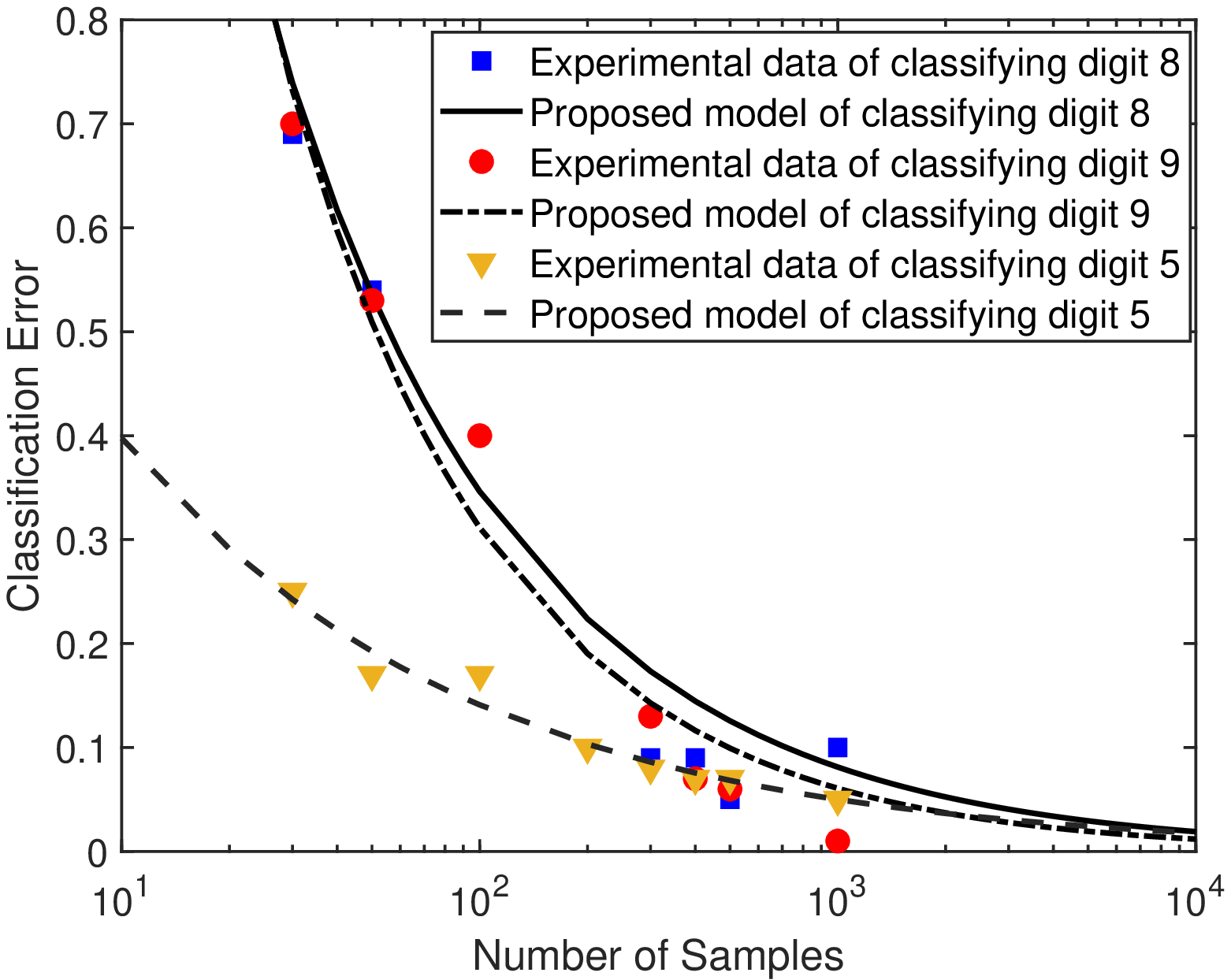}}
  \caption{
  (a) Comparison between the experimental data and the nonlinear classification error model. The parameters in the models are given by
$(a_k,b_k)=(9.27,0.74)$ for CNN and $(a_k,b_k)=(6.94,0.8)$ for SVM;
  (b) Fitting the error function to small datasets. The parameters in the models are given by $(a_k,b_k)=(7.3,0.69)$ for CNN and $(a_k,b_k)=(5.2,0.72)$ for SVM;
  (c) Comparison between different classification tasks.
}
  \label{fig:subfig} 
\end{figure*}

\subsection{Practical Implementation}

One may wonder how could one obtain the fitted model before the actual machine learning model is being trained. There are two ways to address this issue.

\textbf{1) Extrapolation.} More specifically, the error function can be obtained by training the machine learning model on an initial dataset (i.e., with a maximum size of $A_k$) at the edge, and the performance on a future larger dataset can be predicted, such that the edge can decide how many samples to be further collected.
This is called extrapolation \cite{model1}.
For example, by fitting the error function to the first half experimental data of CNN in Fig. 2b (i.e., $v_k=(100,150,200,300)$), we can obtain $(a_k,b_k)=(7.3,0.69)$, and the resultant curve predicts the errors at $v_k=(500,1000,5000,1000)$ very well as shown in Fig.~2b.
Similarly, with $(a_k,b_k)=(5.2,0.72)$ and experimental data of $v_k=(30,50,100,200)$, the proposed model for SVM matches the learning errors at $v_k=(300,400,500,1000)$.
It can be seen that the fitting performance in Fig.~2b is slightly worse than that in Fig. 2a, as we use smaller number of pilot data.
But since our goal is to distinguish different tasks rather than accurate prediction of the learning errors, such an extrapolation method can still guide the resource allocation at the edge.

\textbf{2) Approximation.} This means that we can pre-train a large number of commonly-used models offline (not at the edge) and store their corresponding parameters of $(a_k,b_k)$ in a look-up table at the edge.
Then by choosing a set of parameters from the table, the stored error model can be used to \emph{approximate} the unknown error model at the edge \cite{model2}.
This is because the error functions can share the same trend for two similar tasks, e.g., classifying digit `$8$' and `$9$' with SVM as shown in Fig.~2c.
Notice that there may be a mismatch between the pre-training task and the real task at the edge.
This is the case between classifying digit `$8$' and `$5$' in Fig.~2c.
As a result, it is necessary to carefully measure the similarity between two tasks before choosing the parameters.

\section{The Proposed LCPA Algorithm}

Based on the results in Section III, we can directly approximate the true error function $\Psi_k$ by $\Theta_k$.
However, to account for the approximation error between $\Psi_k$ and $\Theta_k$ (e.g., due to noise in samples or slight mismatch between data used for training and data observed in MTC devices), a weighting factor $\rho_k\geq 1$ can be applied to $\Theta_k$,
where a higher value of $\rho_k$ accounts for a larger approximation error.
Then by replacing $\Psi_k$ with $\rho_k\Theta_k$ and putting \eqref{P0b} into $\Theta_k(v_k,|a_k,b_k)$ to eliminate $\mathbf{v}$, problem $\mathrm{P}$ becomes
\begin{align}
\mathrm{P}1:\mathop{\mathrm{min}}_{\substack{\mathbf{p}}}
\quad&\mathop{\mathrm{max}}_{k=1,\cdots,K}~\rho_k\,\Phi_k(\mathbf{p}),   \nonumber\\
\mathrm{s. t.}\quad&\sum_{k=1}^Kp_k=P_{\mathrm{sum}},\quad p_k\geq 0,\quad \forall k, \label{P1}
\end{align}
where
\begin{align}
&\Phi_k(\mathbf{p})=a_k\left[
\frac{BT}{D_k}\mathrm{log}_2\left(1+\frac{G_{k,k}p_{k}}{\sum_{l\neq k}G_{k,l}p_{l}+
\sigma^2} \right)+A_k
\right]^{-b_k}. \nonumber
\end{align}

To proceed to solve $\mathrm{P}1$, we propose the LCPA algorithm under the framework of MM \cite{mm1}, which constructs a sequence of upper bounds $\{\widetilde{\Phi}_k\}$ on $\{\Phi_k\}$ and replaces $\{\Phi_k\}$ in \eqref{P1} with $\{\widetilde{\Phi}_k\}$ to obtain the surrogate problems.
More specifically, given any feasible solution $\mathbf{p}^\star$ to $\mathrm{P}1$, we define surrogate functions
\begin{align}
\widetilde{\Phi}_{k}(\mathbf{p}|\mathbf{p}^\star)
&=a_k\Bigg\{ \frac{BT}{D_k\mathrm{ln}2\,}
\Bigg[
\mathrm{ln}\left(\sum_{l=1}^KG_{k,l}p_{l}+
\sigma^2 \right)
\nonumber\\
&\quad
{}
-\frac{\sum_{l=1,l\neq k}^KG_{k,l}p_{l}+
\sigma^2}{\sum_{l=1,l\neq k}^KG_{k,l}p^\star_{l}+
\sigma^2} \nonumber
\\&\quad
{}-
\mathrm{ln}\left(\sum_{l=1,l\neq k}^KG_{k,l}p^\star_{l}+
\sigma^2 \right)+1
\Bigg]+A_k
\Bigg\}^{-b_k}. \nonumber
\end{align}
It can be shown that the functions satisfy the following conditions:
\begin{itemize}
\item[(i)] Upper bound condition: $\widetilde{\Phi}_{k}(\mathbf{p}|\mathbf{p}^\star)\geq \Phi_{k}(\mathbf{p})$;

\item[(ii)] Convexity: $\widetilde{\Phi}_k(\mathbf{p}|\mathbf{p}^{\star})$ is convex in $\bm{\mathbf{p}}$.

\item[(iii)] Local equality condition: $\widetilde{\Phi}_{k}(\mathbf{p}^\star|\mathbf{p}^\star)=\Phi_{k}(\mathbf{p}^\star)$ and $\nabla_{\mathbf{p}}\widetilde{\Phi}_{k}(\mathbf{p}^\star|\mathbf{p}^\star)=\nabla_{\mathbf{p}}\Phi_{k}(\mathbf{p}^\star)$.

\end{itemize}

With (i), an upper bound can be directly obtained if we replace the functions $\{\Phi_m\}$ by $\widetilde{\Phi}_m$ around a feasible point.
However, a tighter upper bound can be achieved if we treat the obtained solution as another feasible point and continue to construct the next-round surrogate function.
In particular, assuming that the solution at the $n^{\mathrm{th}}$ iteration is given by $\mathbf{p}^{[n]}$, the following update is executed at the $(n+1)^{\mathrm{th}}$ iteration:
\begin{align}
\mathbf{p}^{[n+1]}=\mathop{\mathrm{argmin}}_{\substack{\mathbf{p}}}
\quad&\mathop{\mathrm{max}}_{k=1,\cdots,K}~\rho_k\,\widetilde{\Phi}_k(\mathbf{p}|\mathbf{p}^{[n]})\nonumber\\
\mathrm{s. t.}\quad~&\sum_{k=1}^Kp_k=P_{\mathrm{sum}},\quad p_k\geq 0,\quad\forall k. \label{P1[n+1]}
\end{align}

Based on (ii), the problem \eqref{P1[n+1]} is convex and can be solved by off-the-shelf software packages (e.g., CVX Mosek \cite{opt1}) for convex programming.
Furthermore, according to (iii) and \cite{mm1}, the sequence
$(\mathbf{p}^{[0]},\mathbf{p}^{[1]},\cdots)$ converges to the KKT solution to $\mathrm{P}1$ for any feasible starting point $\mathbf{p}^{[0]}$ (e.g., we set $\mathbf{p}^{[0]}=P/K\,\bm{1}_K$).
The worst-case complexity for solving $\mathrm{P}1$ is
$\mathcal{O}\left(K^{3.5}\right)$.
The overall architecture of LCPA is shown in Fig.~3a.

\section{Scaling Law of LCPA}

In this section, we investigate the asymptotic case when the number of antennas at the edge approaches infinite (i.e., $N \to +\infty$), which could reveal some insights into LCPA.

As $N\to+\infty$, the channels from different users to the edge would be asymptotically orthogonal and we have
\begin{align}
&G_{k,l}=\frac{|\mathbf{h}_k^H\mathbf{h}_l|^2}{||\mathbf{h}_k||_2^2}\to 0,\quad\forall k\neq l.
\end{align}
Based on such orthogonality feature, and putting $G_{k,l}=0$ for $k\neq l$ into $\Phi_k$ in $\rm{P}1$, the problem $\mathrm{P}1$ under $N\to +\infty$ is equivalent to
\begin{subequations}
\begin{align}
\mathrm{P}2:\mathop{\mathrm{min}}_{\substack{\mathbf{p},\,\mu}}
\quad&\mu,
\nonumber\\
\quad\quad\quad~\mathrm{s. t.}\quad~&
\rho_ka_k\left(
\frac{BT}{D_k}\mathrm{log}_2\left(1+\frac{G_{k,k}p_{k}}{\sigma^2} \right)+A_k
\right)^{-b_k}
\nonumber\\
&
\leq\mu,\quad\forall k,
\label{P3a}\\
&
\sum_{k=1}^Kp_k= P_{\mathrm{sum}},\quad p_k\geq 0,\quad\forall k,
\label{P3b}
\end{align}
\end{subequations}
where $\mu\in[0,1]$ is a slack variable and has the interpretation of \emph{classification error level}.
The following proposition gives the optimal solution to $\mathrm{P}2$ (proved based on the Karush-Kuhn-Tucker conditions of $\mathrm{P}2$; See \cite{edge2} for more details).
\begin{proposition}
The optimal $\mathbf{p}^*$ to $\mathrm{P}2$ is
\begin{align}\label{pk*}
p_k^*(\mu^*)=&\Bigg[\frac{\sigma^2}{G_{k,k}}\,\mathrm{exp}\left(\frac{D_k\mathrm{ln}2\,}{BT}\left[\left(\frac{\mu^*}{\rho_ka_k}\right)^{-1/b_k}-
A_k\right]\right)
\nonumber\\
&
-\frac{\sigma^2}{G_{k,k}}\Bigg]^+,\quad k=1,\cdots,K,
\end{align}
where $\mu^*$ satisfies $\sum_{k=1}^Kp_k^*(\mu^*)=P_{\mathrm{sum}}$.
\end{proposition}

To efficiently compute the classification error level $\mu^*$, it is observed that the function $p_k^*(\mu^*)$ is a decreasing function of $\mu^*$.
Therefore, the $\mu^*$ can be obtained from solving $\sum_{k=1}^Kp_k^*(\mu^*)=P_{\mathrm{sum}}$ using bisection method within interval $[0,1]$.
The bisection method has a complexity of $\mathcal{O}(\mathrm{log}\left(\frac{1}{\epsilon}\right)K)$.

According to \textbf{Proposition 1}, the user transmit power $p_k$ is inversely proportional to the wireless channel gain $G_{k,k}=||\mathbf{h}_k||_2^2$.
However, it is exponentially dependent on the classification error level $\mu$ and the error model parameters $(a_k,b_k,D_k,A_k)$.
Moreover, among all parameters, $b_k$, which is related to the complexity of the dataset, is the most important factor, since $b_k$ is involved in both the power and exponential functions.
The above observations disclose that in edge intelligence systems, the classification error model parameters will have more significant impacts on the physical-layer design than those of the wireless channels.
Therefore, it is important to conduct sensitivity analysis of the power allocation solution with respect to the accuracy in estimating the parameters $(a_k, b_k)$ of the classification error model.

\section{Simulation Results and Discussions}

This section provides simulation results to evaluate the performance of the proposed algorithms.
It is assumed that the noise power $\sigma^2=-87\,\mathrm{dBm}$  (corresponding to power spectral density $-140\,\mathrm{dBm/Hz}$ with $180\,\mathrm{kHz}$ bandwidth \cite{iot3}), which includes thermal noise and receiver noise.
Unless otherwise specified, the total transmit power at users is set to $P_{\mathrm{sum}}=13\,\mathrm{dBm}$ (i.e., $20\,\rm{mW}$), with the time budget $T=5\,\mathrm{s}$ and the communication bandwidth $B=180\,\mathrm{kHz}$.
The path loss of the user $k$ $\varrho_{k}=-100\,\rm{dB}$ is adopted, and $\mathbf{h}_{k}$ is generated according to $\mathcal{CN}(\mathbf{0},\varrho_{k}\mathbf{I}_N)$.
Each point in the figures is obtained by averaging over $10$ simulation runs, with independent channels in each run.
All optimization problems are solved by Matlab R2015b on a desktop with Intel Core i5-4570 CPU at 3.2\,GHz and 8\,GB RAM.
All the classifiers are trained by Python 3.6 on a GPU server with Intel Core i7-6800 CPU at 3.4\,GHz and GeForce GTX 1080 GPU.

For the edge intelligence system, we consider the the case of $K=2$ with aforementioned CNN and SVM classifiers at the edge: i) Classification of MNIST dataset \cite{MNIST} via deep CNN; ii) Classification of digits dataset in Scikit-learn \cite{sklearn} via SVM.
The data amount of each sample is $D_1=6276\,\mathrm{bits}$ for MNIST dataset and $D_2=324\,\mathrm{bits}$ for digits dataset in Scikit-learn.
It is assumed that there are $A_1=300$ CNN samples and $A_2=200$ SVM samples before transmission.
The parameters in the two error models are obtained by fitting the model \eqref{model1} to the initial datasets at the edge, and they are given by $(a_1,b_1)=(7.3,0.69)$ for CNN and $(a_2,b_2)=(5.2,0.72)$ for SVM as in Fig.~2b.
Finally, it is assumed that $(\rho_1,\rho_2)=(1,1.2)$ since the approximation error of SVM in Fig.~2b is larger than that of CNN.

\begin{figure*}[htbp]
 \centering
      \subfigure[]{
    \label{fig:subfig:b} 
    \includegraphics[height=33mm]{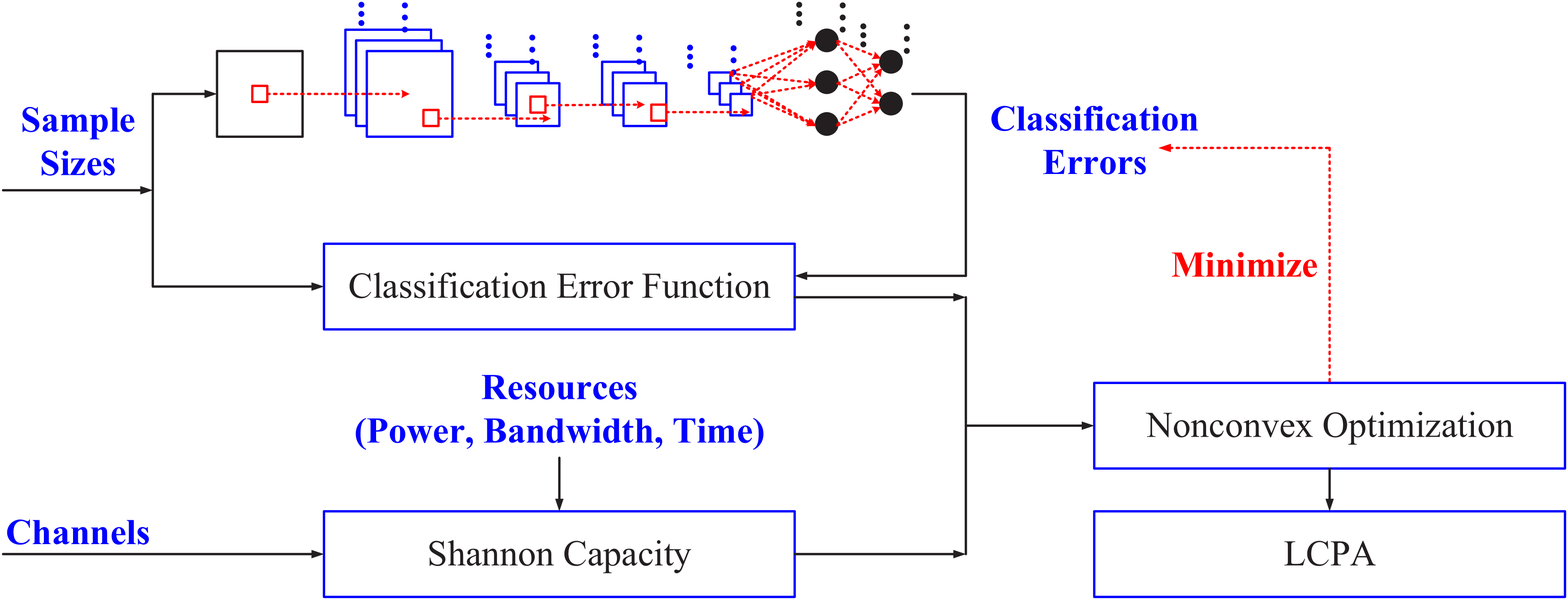}}
  \subfigure[]{
    \label{fig:subfig:b} 
    \includegraphics[height=33mm]{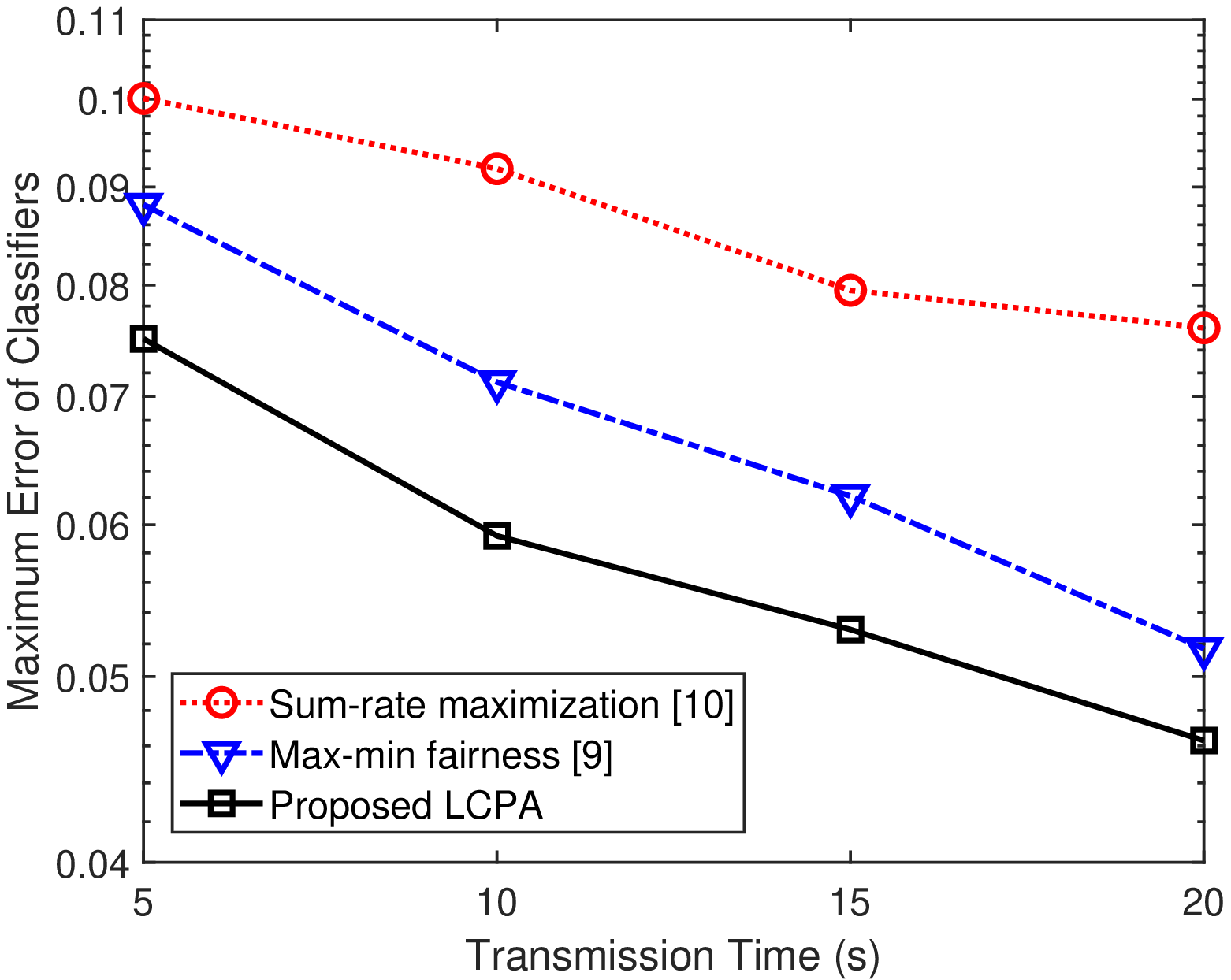}}
      \subfigure[]{
    \label{fig:subfig:b} 
    \includegraphics[height=33mm]{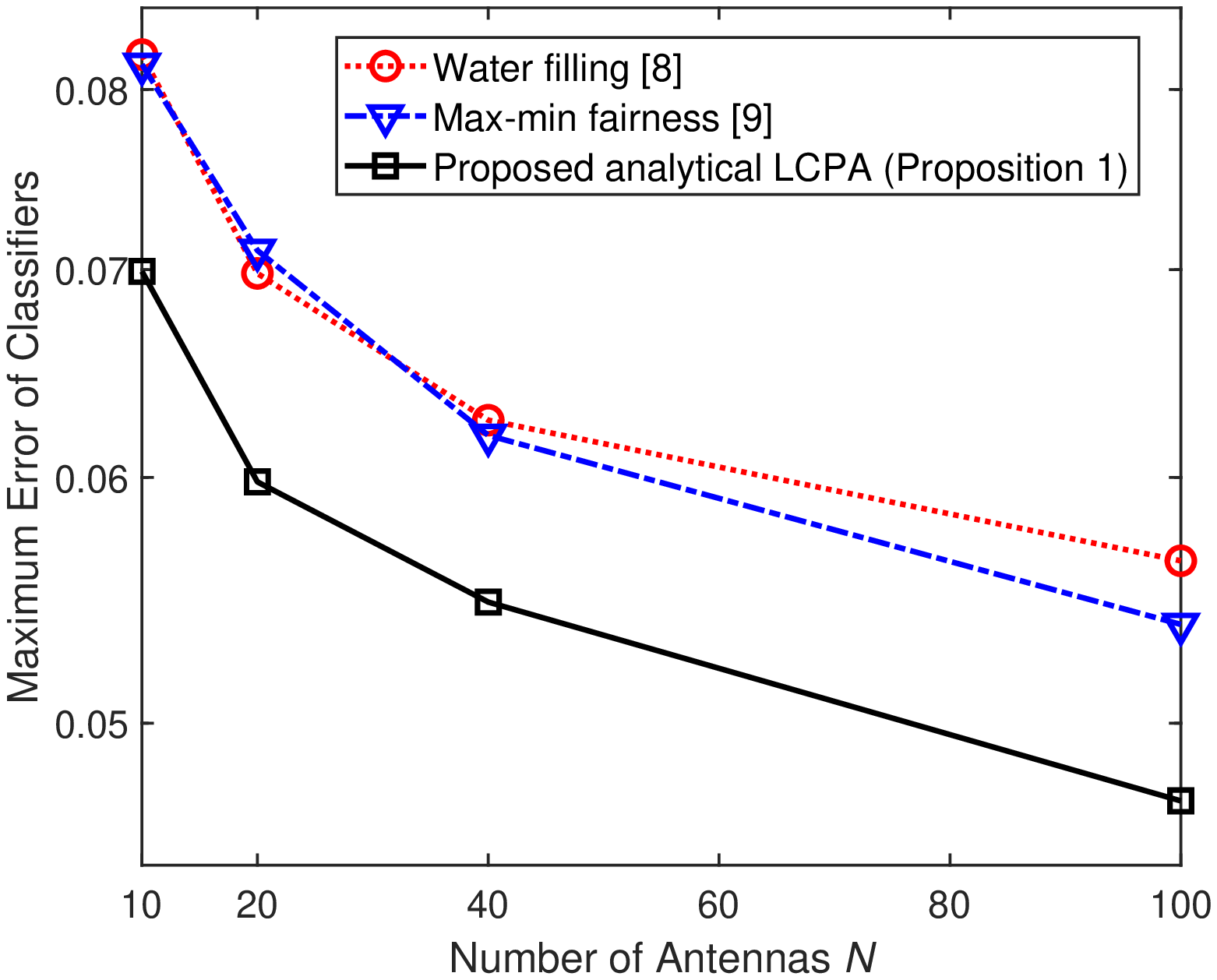}}
  \caption{
    (a) Architecture of edge intelligence with LCPA; (b) Maximum error of classifiers versus total transmission time $T$ in $\mathrm{s}$ when $K=2$ and $N=4$; (c) Maximum error of classifiers versus number of antennas $N$ when $K=2$.
}
  \label{fig:subfig} 
\end{figure*}

To begin with, the case of $N=4$ and $K=2$ is simulated.
Under the above settings, the classification error (obtained from the machine learning experiment using the sample sizes from the power allocation algorithms) versus the total transmission time $T$ in $\mathrm{s}$ is shown in Fig. 3b.
Besides the proposed Algorithm~1, we also simulate two benchmark schemes: 1) Max-min fairness scheme \cite[Sec. II-C]{fair}, which computes the dominate eigenvector of the ``extended uplink coupling matrix''; 2) Sum-rate maximization scheme \cite[Sec. IV]{sumrate}, which uses difference of convex programming to allocate power.
It can be seen from Fig.~3b that the proposed LCPA algorithm with $10$ iterations significantly reduces the classification error compared to other schemes, and the gap concisely quantifies the benefit brought by more training images for CNN under joint communication and learning design.
For example, at $T=20\,\mathrm{s}$ in Fig.~3b, the proposed LCPA collects $1604$ MNIST images on average, while the sum-rate maximization and the max-min fairness schemes obtain $1036$ images and $1148$ images, respectively.

\begin{table}[!h]
\caption{Comparison of Average Transmit Power in $\mathrm{mW}$ When $N=10$} 
\centering 
\begin{tabular}{|c|c|c|c|c|}
\hline
User & Analytical LCPA & Water-filling & Max-min fairness \\
\hline
$k=1$ (CNN) & $19.8476$ & $9.9862$ & $10.0869$ \\
\hline
$k=2$ (SVM) & $ 0.1524$ & $10.0138$ & $9.9131$ \\
\hline
\end{tabular}
\end{table}

To get more insight into the edge intelligence system, the classification error versus the number of antennas $N=\{10,20,40,100\}$ with $K=2$ is shown in Fig.~3c.
It can be seen from Fig.~3c that the classification error decreases as the number of antennas increases, which demonstrates the advantage of employing massive MIMO at the edge.
More importantly, the proposed analytical solution in \textbf{Proposition 1} outperforms the water-filling and max-min fairness schemes even at a relatively small number of antennas $N=10$.
This is achieved by allocating much more power resources to the first MTC user (i.e., the user uploading MNIST dataset) as shown in Table~I, because training CNN is more difficult than training SVM.

\section{Conclusions}

This paper has studied the LCPA at the edge.
By adopting an empirical classification error model, efficient edge resource allocation has been obtained via the LCPA algorithm.
Based on asymptotic analysis, the scaling law of learning centric communication has been revealed.
Simulation results have shown that the proposed LCPA algorithm achieves lower prediction errors than all traditional power allocation schemes.

\section{Acknowledgement}
The work was supported by the National Natural Science Foundation of China under Grants 61771232 and 61773197, the Shenzhen Basic Research Project under Grant JCYJ20190809142403596, and the Natural Science Foundation of Guangdong Province under Grant 2017A030313335.
This work was also supported by the U.S. National Science Foundation under Grants CCF-0939370, CCF-1513915 and CCF-1908308.

\end{document}